\documentclass[journal=jacsat,manuscript=article]{achemso}

\author{Pim Witte}
\affiliation{Debye Institute for Nanomaterials Science, Utrecht University, 3508 TA Utrecht, the Netherlands.}
\author{Denis Sheptyakov}
\affiliation{Laboratory for Neutron Scattering and Imaging, Paul Scherrer Institut, Forschungsstrasse 111, Villigen CH-5232, Switzerland.}
\author{Elsa Lhotel}
\affiliation{Institut Néel, CNRS, Université Grenoble Alpes, 38042 Grenoble, France.}
\author{Nongnuch Artrith}
\affiliation{Debye Institute for Nanomaterials Science, Utrecht University, 3508 TA Utrecht, the Netherlands.}
\author{Robin de Hoogh}
\affiliation{Debye Institute for Nanomaterials Science, Utrecht University, 3508 TA Utrecht, the Netherlands.}
\author{Giuditta Perversi}
\affiliation{Maastricht Science Programme, Faculty of Science and Engineering, Maastricht University, Paul Henri Spaaklaan 1, 6229 EN Maastricht, the Netherlands.}
\author{Kim Lefmann}
\affiliation{Nanoscience Center, Niels Bohr Institute, University of Copenhagen, 2100 Copenhagen Ø, Denmark.}
\author{Machteld E. Kamminga}
\email{m.e.kamminga@uu.nl} 
\affiliation{Debye Institute for Nanomaterials Science, Utrecht University, 3508 TA Utrecht, the Netherlands.}

\usepackage{graphicx} 
\usepackage{dcolumn}
\usepackage{bm}
\usepackage[version=4]{mhchem}
\usepackage{lipsum}
\usepackage{gensymb} 
\usepackage{microtype} 
\usepackage{amsmath}
\usepackage{blindtext}
\usepackage{amsfonts}
\usepackage{hyperref}
\usepackage{upgreek}
\usepackage{mathrsfs}
\usepackage{braket}
\usepackage{nicefrac, xfrac}
\usepackage{anyfontsize}
\usepackage{float}
\usepackage{color}
\usepackage{multirow}
\usepackage{multicol}

\title{
Complex magnetic interactions in geometrically frustrated TbOF
}

\begin{document}

\begin{abstract}
  \noindent 
We have identified TbOF as a unique frustrated and mixed-anion lattice, hosting unconventional magnetism. By means of magnetization, specific heat and neutron diffraction measurements down to 90 mK, as well as DFT calculations, we present a comprehensive study of the magnetic and structural properties of \ce{TbOF}. We show that at 9.7 K, TbOF undergoes a structural phase transition accompanied by short-range magnetic correlations, in contrast to previously proposed long-range antiferromagnetic order. At lower temperatures, we observe two magnetic ordering transitions, consisting of incommensurate spin density waves and antiferromagnetic and ferromagnetic correlations. Furthermore, we observe metastable and hysteresis behavior below 2.0 K, highlighting the richness of complex magnetic interactions in TbOF. These results uniquely clarify the magnetic phase diagram of \ce{TbOF} and highlight the intricate interplay between structure and magnetism in rare-earth oxyfluorides.
\end{abstract}

\section{Introduction}

Lanthanide (\ce{\textit{Ln}^{3+}}) cations have proven to be essential components of luminescent materials, as they enable applications ranging from energy-efficient lighting and high-resolution displays, to bio-imaging and optical telecommunication \cite{Tessitore2023LanthanideLuminescence, Alexander2025, Zhu2023LanthanideBiologicalImaging, Li2025}. In this context, lanthanide oxyfluorides (\textit{Ln}OF) are an especially attractive family. As mixed-anion materials, oxyfluorides combine the low phonon energies characteristic of rare-earth fluorides that preserve radiative transitions, with chemical stability associated with oxides, preserving both optical performance and structural integrity \cite{Yi2011, Zhang2014, Ngo2021, He2016}. It is for these reasons that oxyfluorides have been considered one of the most successful host lattices in phosphors.\cite{Wang1993,Shen2004,Yu2009}

Notably, the same $4f$ electrons that produce the long emission lifetimes and host sharp and insensitive optical transitions that make these ions so suitable for optical applications, also endow materials with well-localised magnetic moments that can produce rich magnetic behavior \cite{Coey2020}. Due to the strong anisotropy and weak magnetic coupling of $4f$ ions, lanthanide based compounds provide a fertile ground for unconventional magnetism. In frustrated geometries, \textit{e.g.} in the corner- and edge-sharing \ce{\textit{Ln}^{3+}} networks found in pyrochlores and related structures, spin-ice and spin-liquid-behavior can be realized \cite{Bramwell2001, Castelnovo2008, Fennell2009, Lhotel2015, Clark2019}.

However, the magnetic behaviour induced by the \ce{\textit{Ln}^{3+}} ions in the \textit{Ln}OF lattice has received much less attention, despite the fact that the \ce{\textit{Ln}^{3+}} ions in rhombohedral \textit{Ln}OF compounds (\textit{Ln} = Nd, Eu, Gd, Tb, Dy, Ho, Er) form a network of face-sharing tetrahedra, which could be seen as a three-dimensional analogue to geometrically frustrated triangular lattices that are known to give rise to exotic magnetic ground states. The coexistence of oxygen and fluorine in these mixed anion compounds also introduces distinct \textit{Ln}–O and \textit{Ln}–F bonds, providing anisotropic exchange pathways. Rhombohedral oxyfluorides therefore could potentially host magnetic frustration and emergent behavior in addition to quantum fluctuations \cite{Clark2019, Ji2021}. 

NdOF, HoOF, and ErOF have all been reported to remain paramagnetic down to cryogenic temperatures, consistent with the strong frustration and suppressed magnetic ordering, expected in structures like these \cite{Dutton2012, Ji2021, Beaury1998NdOF, Beaury2002NdOF}. Surprisingly, \ce{TbOF} is an intriguing outlier amongst the rhombohedral oxyfluorides: Dutton \textit{et al.} reported that \ce{TbOF} undergoes a transition near $T_{\text{m}} = 10$ K into an antiferromagnetic-like ordered state, accompanied by a field-induced metamagnetic transition at $\mu_0 H_{\text{t}} = 1.8$ T at 2 K \cite{Dutton2012}. Not only does this make \ce{TbOF} one of the only oxyfluorides suggested to achieve long-range magnetic order, it also hints towards complex magnetic behavior in this material system. Despite this, both the magnetic structure and the nature of the 10 K transition have remained unresolved. To understand how the frustration present in other oxyfluorides is overcome, it is crucial that the magnetic ground state is probed.

In this work we investigate the magnetic behavior of \ce{TbOF} in detail using a combination of magnetometry, specific heat measurements, and neutron diffraction down to 90 mK. We demonstrate that the reported peak in magnetization near 10 K does not correspond to any conventional antiferromagnetic ordering, but rather marks the onset of short-range magnetic correlations concurrent with a structural distortion of the rhombohedral lattice. Upon further cooling, the system develops incommensurate, long-range order that evolves at lower temperatures. These results establish \ce{TbOF} as a rare example of a lanthanide oxyfluoride in which the structural and magnetic degrees of freedom interplay to overcome geometric frustration and generate unconventional, temperature-dependent magnetic phases. This rich magnetic phase diagram provides insight into structure-property relationships in magnetic $4f$ materials.

\section{Experimental Methods}
\noindent Polycrystalline \ce{TbOF} was synthesized by a conventional solid-state reaction following established procedures \cite{Beaury2002NdOF, Beaury1998NdOF, Shinn1969Oxyfluorides, Baenziger1954UnitCell}. Stoichiometric amounts of \ce{Tb4O7} ($99.99\%$ Rhône-Poulenc) and \ce{NH_4F} ($\geq98\%$ Sigma-Aldrich) were mixed in a 1:6 molar ratio and thoroughly ground to homogeneity. The resulting mixture was transferred to an alumina crucible, heated to 900 \degree C for 1 h, and subsequently cooled to room temperature. The reaction product, \ce{TbOF}, was obtained as a white powder. Phase purity was confirmed by powder X-ray diffraction, and data for structure refinement were collected at room temperature using a PANalytical AERIS diffractometer (Cu K\(\alpha\) radiation, Bragg-Brentano geometry, 10–90\degree \ \(2\theta\) range). Refinement of the crystal structure was conducted using the JANA2020 software \cite{Petricek2023Jana}.

Our DFT calculations are based on the plane-wave pseudopotential code Vienna Ab initio Simulation Package (VASP) \cite{Kresse1996, Kresse1996b}, using the noncollinear version of VASP 6.4. Exchange-correlation effects were described with the $\text{r}^2$SCAN meta-generalized gradient approximation \cite{Furness2020}, with the projector augmented wave basis \cite{Blochl1994, Kresse1999}. The localized Tb $4f$ electrons were treated using the rotationally invariant DFT+U method \cite{Burnett2024}. A plane-wave energy cutoff of 400 eV was employed, with spin polarization included. The electronic self-consistent field (SCF) convergence criteria were set to $10^{-4}$ eV for energy and 0.01 eV/Å for forces, and all calculations were carried out until full electronic convergence was achieved. Further details are provided in the SI.

DC magnetization measurements were performed using a MPMS XL 5T Quantum Design magnetometer in the 1.8 K – 300 K temperature range and 0 – 5 T field range, following both zero-field-cooled (ZFC) and field-cooled (FC) procedures. Additional DC magnetization measurements below 4 K were performed using a SQUID magnetometer equipped with a dilution refrigerator developed at the Institut Néel CNRS in the temperature range of 100 mK to 4.2 K in both ZFC and FC procedures. The sample was mixed with Apiezon grease in a copper pouch to ensure a good thermalization.

Specific heat measurements were performed by the standard $2\uptau$ method using a Quantum Design Physical Property Measurement System (PPMS) calorimeter equipped with a $^{3}\text{He}$ pumping system, to extend the temperature range down to 350 mK.

Neutron powder diffraction experiments were performed on the High Resolution Powder Diffractometer for Thermal Neutrons (HRPT) \cite{Fischer2000HRPT} at the Swiss Spallation Neutron Source (SINQ), at the Paul Scherrer Institute in Switzerland. A polycrystalline sample of \ce{TbOF} with a total mass of 2.5 g  was loaded in vanadium cans (6 and 8 mm inner diameter) and measured down to 1.55 K using an Orange He-cryostat ($\lambda = 1.886$ \AA). In addition, the same sample was loaded in a copper can (6 mm inner diameter), and measured down to 90 mK using a dilution insert ($\lambda = 1.494$ \AA, 1.886 \AA \ and 2.445 \AA). Refinement of the nuclear and magnetic structures were carried out using the FULLPROF program suite \cite{RodrguezCarvajal1993FULLPROF}. The magnetic propagation vectors were determined using the K-search utility implemented in the FullProf suite. Representational analysis was carried out using the BasIreps program.\cite{basireps} The crystal structures are drawn using VESTA \cite{Momma2011VESTA}.

\section{Results}
\begin{figure}[t]
    \centering
    \includegraphics{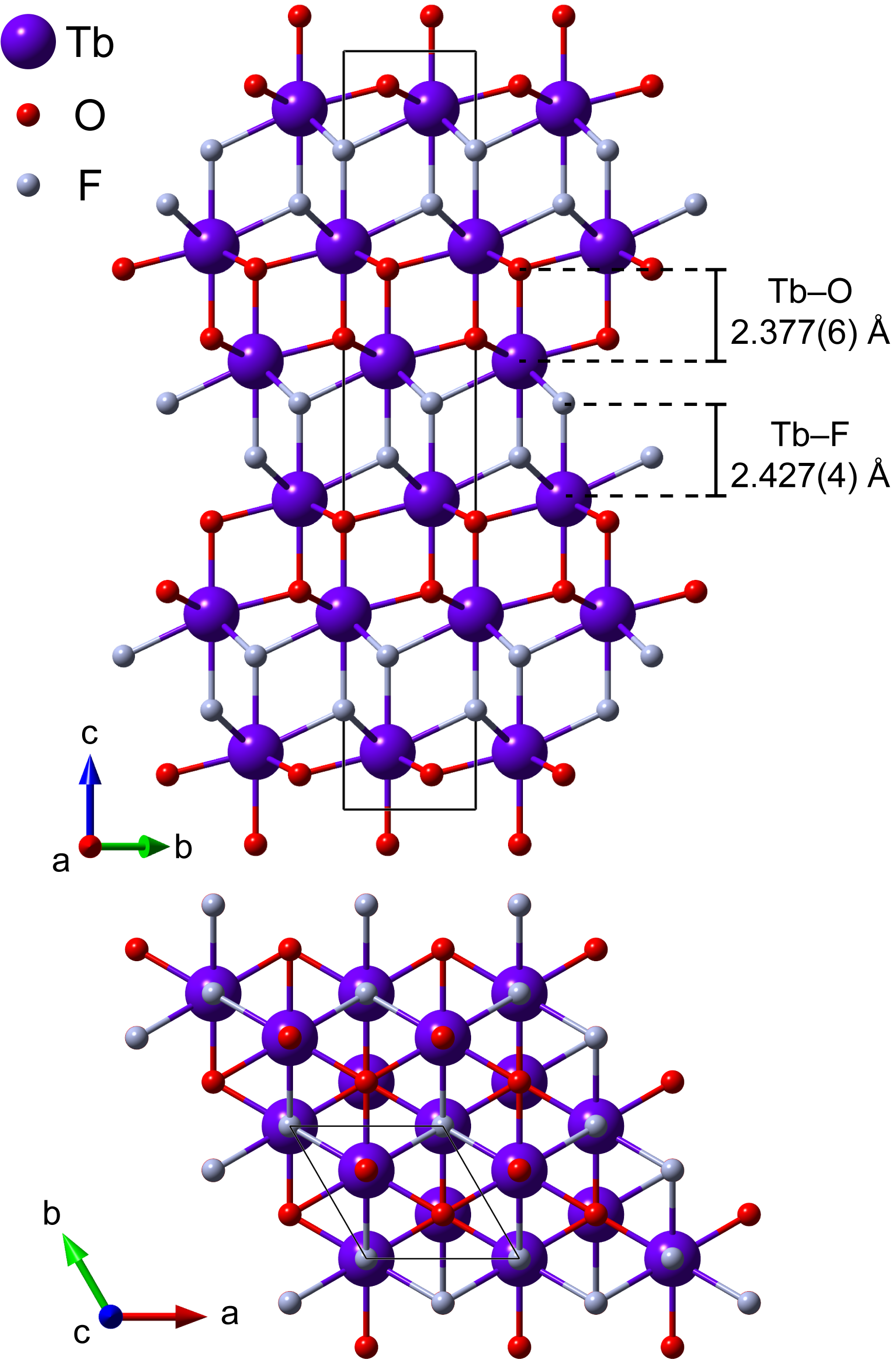}
    \caption{Crystal structure of TbOF at room temperature, depicted along the \textit{a}-axis (top) and \textit{c}-axis (bottom). Relevant bond lengths are indicated.}
    \label{fig:CrystalStructure}
\end{figure}

Using our solid-state synthesis method, we have obtained phase-pure \ce{TbOF} powder, as demonstrated by the Rietveld refinement to the powder X-ray diffraction data shown in Figure S1 in the Supporting Information (SI). The crystal structure adopts the  $\text{R}\bar{3}m$ space group. As shown in Figure \ref{fig:CrystalStructure}, each Tb atom coordinates with four O atoms and four F atoms, forming triangular layers of Tb atoms. These Tb layers are stacked along the \textit{c}-direction, with a (1/3, 1/3) offset in the \textit{ab}-plane, and alternately separated by O and F layers. Given the relative positions of the magnetic Tb atoms and their triangular framework, \ce{TbOF} is a clear template for observing magnetic frustration.

The rhombohedral $\text{R}\bar{3}m$ structure adopted by \ce{TbOF} and other rare-earth oxyfluorides contains distinct short and long lanthanide–anion bonds, with oxide and fluorine ions. However, neither X-ray nor neutron diffraction experiments can distinguish well between oxygen and fluorine atoms. Consequently, we noticed both configurations in the Inorganic Crystal Structure Database (ICSD).\cite{Zagorac2019} While most entries represent \textit{Ln}OF compounds wherein the \textit{Ln}-O bonds are significantly shorter than the \textit{Ln}-F bonds, the reverse is listed for LaOF \cite{Zachariasen1951}. However, in order to understand the magnetic behavior of \ce{TbOF} in detail, it is important to get a clear understanding of the exchange pathways. Therefore, we deem it crucial to clarify which ion takes which crystallographic site in TbOF. An initial study by Templeton in 1957, performing electrostatic calculations on YOF, indicated that the Y-O bonds are the shorter bonds.\cite{Templeton1957} Here we perform DFT calculations to assign the O and F positions in the refined \ce{TbOF} structure. These calculations found that the \ce{TbOF} structure is energetically and dynamically more stable in the case where the Tb–F bonds are indeed longer than the Tb–O bonds, in agreement with Templeton. Further details regarding our calculations and the lattice parameters used are written in the SI and Table S1. Moreover, these findings are in agreement with bond valence sum calculations performed on both configurations, as listed in Table S2.

\begin{figure}[t]
    \centering
    \includegraphics{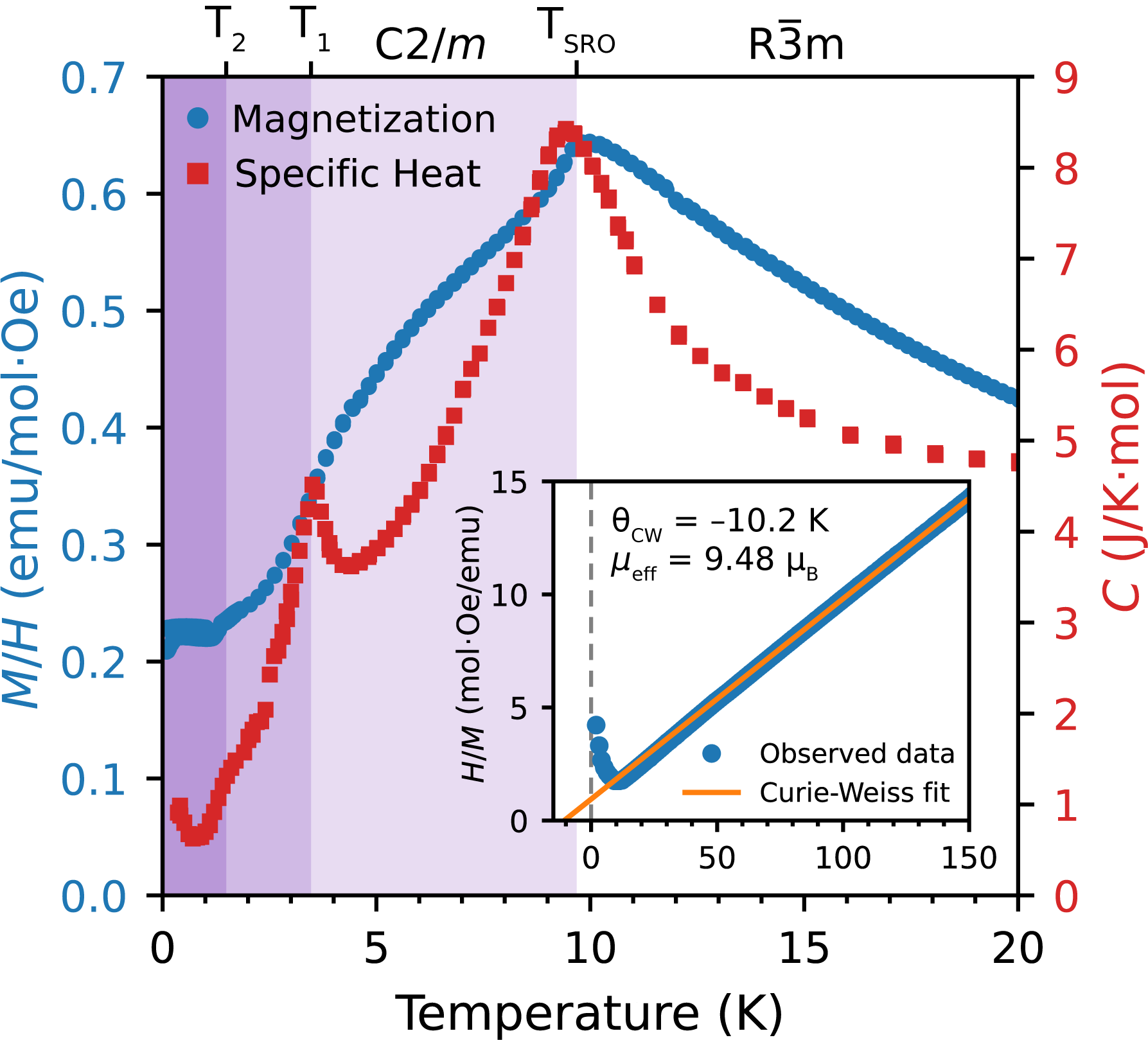}
    \caption{Temperature-dependence of the magnetization ($M/H$), blue circles, left axis) and specific heat (\textit{C}, red squares, right axis) for \ce{TbOF}. The plotted magnetization is a compilation of measurements above and below 1.8 K, measured under applied magnetic fields of 0.1 and 0.01 T, respectively. The specific heat was measured in the absence of an applied magnetic field. $T_1$, $T_2$ and  $T_{\text{RSO}}$, represent the two magnetic order and short-range-order transitions, respectively. The insert displays the inverse of the magnetization with a Curie-Weiss fit plotted on top.}
    \label{fig:SQUID+Specific Heat}
\end{figure}

In order to get insight in the magnetic properties of TbOF, we measured the magnetization as a function of temperature, down to 90 mK. As shown in Figure \ref{fig:SQUID+Specific Heat}, there is a clear correspondence between the specific heat data and the temperature-dependent magnetization, revealing a series of magnetic anomalies at low temperatures. We note that the field-cooled and zero-field-cooled magnetization curves fully overlap in most of the temperature range, with the exception of very low temperature, as discussed below.

In addition to the (relatively broad) peak at 9.7 K, a second, sharper peak appears near 3.5 K in the specific heat data, that coincides with a maximum in $\frac{\text{d}M}{\text{d}T}$ (see Figure S2 in SI). Given the fitted Weiss temperature $\theta$ = -10.2 K, the broad peak at 9.7 K could point towards an antiferromagnetic phase transition. In addition to these observations, which are in agreement with what was previously reported by Dutton \textit{et al.},\cite{Dutton2012} our measurements at low temperatures pinpoint a third transition below 1.6 K in the specific heat data that also coincides with another maximum in $\frac{\text{d}M}{\text{d}T}$ (see Figure S2 in SI), suggesting the possible presence of an additional low-temperature magnetic phase transition. The small upturn observed in the specific heat data is most probably due to the Tb hyperfine contribution as it matches previous reports well \cite{Blte1969, Massalami2003}.

Furthermore, our isothermal magnetization measurements confirm the previously reported field-induced transition at $\mu_0 H=1.8$ T at 2 K,\cite{Dutton2012} indicative of a possible spin-flip transition, as shown in Figure S3 in the SI. This in combination with the successive transitions highlights the rich and complex (magnetic) phase diagram of \ce{TbOF}.

\begin{figure}[t]
    \centering
    \includegraphics{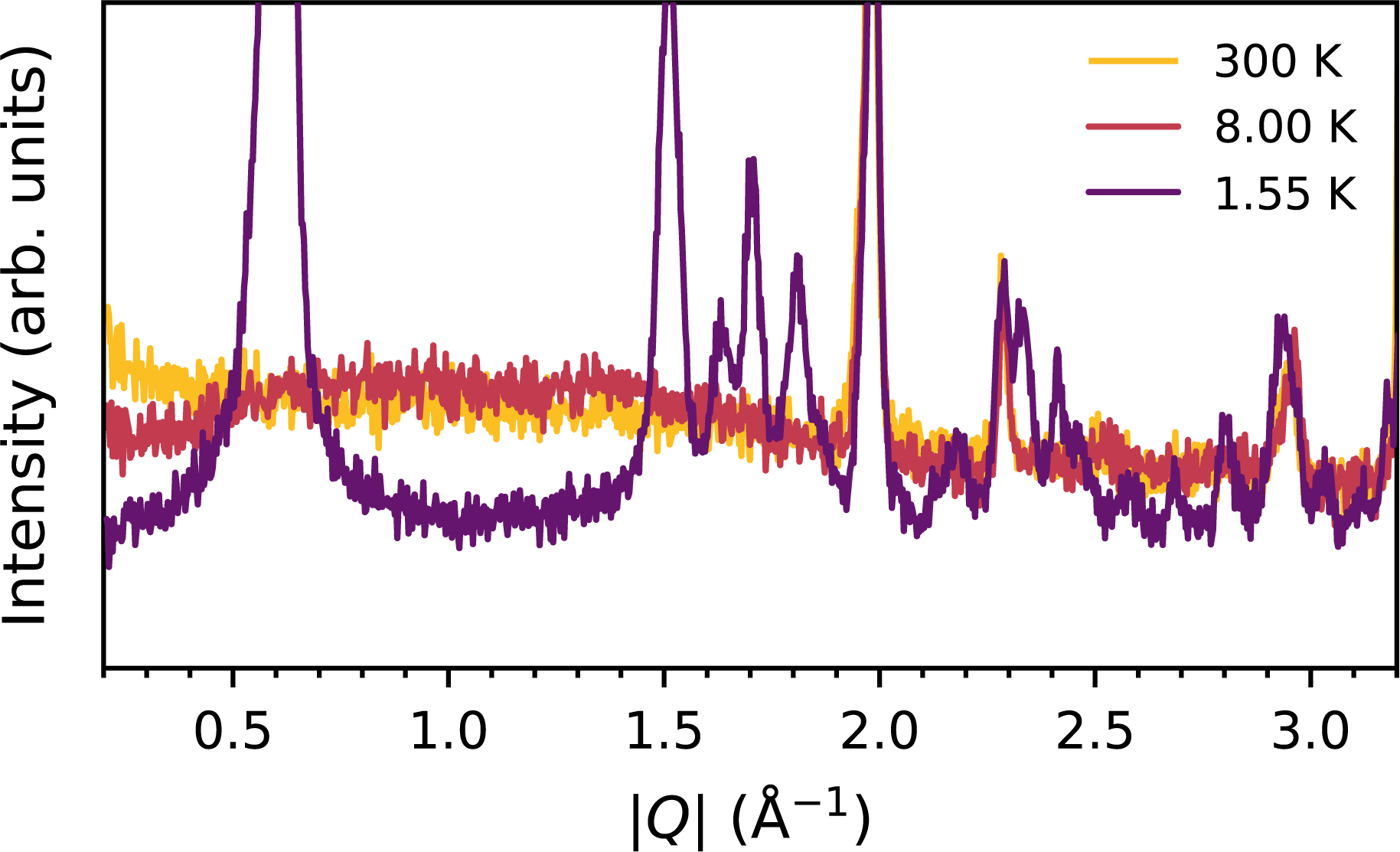}
    \caption{Raw powder neutron diffraction data of \ce{TbOF} ($\lambda = 1.886 \ \text{\AA}$), collected at 300 K, 8.0 K, and 1.6 K. The 300 K pattern (yellow) shows only nuclear reflections consistent with the $\text{R}\bar{3}m$ structure. The 8.0 K dataset (red), recorded below the 9.7 K phase transition shows a reduction in background compared to 300 K, with a concentration of increased scattering in the form of a very broad diffuse maximum centered around 1.0 $\text{Å}^{-1}$, indicative of short-range magnetic correlations. At 1.55 K (purple), below the 3.5 K phase transition, this diffuse feature disappears and magnetic Bragg reflections emerge, consistent with long-range magnetic order.}
    \label{fig:Short-range bump HRPT}
\end{figure}

To elucidate the nature of the magnetic phases, we have performed neutron diffraction measurements down to 90 mK. No evidence for long range magnetic order could be observed at 8 K, indicating that the 9.7 K peak in both the specific heat and magnetization cannot be attributed to the onset of long-range three-dimensional order, as previously proposed \cite{Dutton2012}. Most interestingly, as shown in Figure \ref{fig:Short-range bump HRPT}, some concentration of the scattering in form of a very broad diffuse maximum spanning between $\sim$ 0.4 and $\sim$ 1.7 $\text{Å}^{-1}$ with some decreased scattering in the background regions at both sides of it. This suggests that a part of a paramagnetic diffuse scattering is being shifted into the broad maxima, indicative of short-range magnetic correlations. Note that a more elaborate study on the nature of these short-range magnetic correlations would require other experimental techniques, such as ESR, NMR, $\upmu$SR and inelastic neutron experiments, all of which are beyond the scope of this work. At lower temperatures, below the phase transition at 3.5 K, magnetic Bragg reflections do emerge, indicating the onset of long-range magnetic order at temperatures well below 9.7 K, which is more as expected, as \ce{\textit{Ln}^{3+}} containing lattices tend to order at particularly low temperatures, or not order at all.

Although the 9.7 K transition does not correspond to the onset of long-range magnetic order, we find that it is accompanied by a structural phase transition. This is clear from the peak splitting shown in Figure S4 in the SI. The low background and high resolution of the data obtained from HRPT \cite{Fischer2000HRPT}, allowed us to determine the crystal structure. The diffraction pattern with split reflections was indexed on a monoclinic cell with a symmetry of $\text{C}2/m$ at 8 K, originally a rhombohedral $\text{R}\bar{3}m$ structure at 14 K. See Table \ref{Table: Nuclear phases} for crystallographic parameters. The 8 K structural model has been built based on the assumption of the preserved main structure motif, and refined against the neutron powder data. This monoclinic distortion primarily affects the local tetrahedral geometry by lifting the three-fold symmetry of the basal triangular faces. In the rhombohedral $\text{R}\bar{3}m$ phase, the basal face of the face-sharing Tb tetrahedron is equilateral; upon symmetry lowering to the monoclinic $\text{C}2/m$ phase, these triangles distort into isosceles units with one shortened Tb–Tb distance of 3.82750(6) \AA \ and two longer distances of 3.84080(5) \AA. Moreover, the Tb-O and Tb-F distances along the $c$-axis of the rhombohedral cell change from 2.32514(3) Å to 2.28515(4) Å and 2.41410(3) Å to 2.45628(4) Å, respectively. This means that the shorter Tb-O distances get even shorter and the longer Tb-F distances get even longer. Figure S5 in the SI shows the crystal structures of the rhombohedral and monoclinic phases, highlighting these changes in exchange paths.

\begin{table}
\centering
\caption{Crystallographic parameters for TbOF at 14 and 8 K, obtained from Rietveld refinement of powder neutron diffraction data. Atom positions given in $x$, $y$, $z$ fractional coordinates.}
\begin{tabular}{|l|l|l|}
 \hline
                        & 14 K & 8 K  \\
 \hline \hline
crystal system          & Trigonal    & Monoclinic    \\
space group             & $\text{R}\bar{3}m$    & $\text{C}2/m$    \\
Z                       & 6    & 4    \\
$a$ (Å)                   & 3.8375(1)  & 6.6612(1) \\
$b$ (Å)                   & 3.8375(1)  & 3.8279(1) \\
$c$ (Å)                   & 19.0898(2)  & 6.7431(2) \\
$\alpha$ ($\degree$)    & 90    & 90    \\
$\beta$ ($\degree$)     & 90    & 109.335(5)    \\
$\gamma$ ($\degree$)    & 120    & 90    \\
volume ($\text{Å}^3$)   & 243.474(4)    & 162.172(4)    \\
Tb ($x$,$y$,$z$) & 0.00000    & 0.25900   \\
 & 0.00000    & 0.00000   \\
 & 0.74254    & 0.27550   \\
O ($x$,$y$,$z$) & 0.00000    & 0.37910   \\
 & 0.00000    & 0.00000   \\
 & 0.62074    & 0.63470  \\
 F ($x$,$y$,$z$) & 0.00000    & 0.12940   \\
 & 0.00000    & 0.00000   \\
 & 0.86900    & 0.88940  \\
  \hline
\end{tabular} \label{Table: Nuclear phases}
\end{table}

We propose that this symmetry reduction can be understood as frustration relief from the magnetic lattice, which would mean that the structural transition has a magnetic origin, even if it is not associated with long-range magnetic ordering. This is justified by the following observations: 1) the structural transition is accompanied by the emergence of short-range magnetic correlations, 2) the transition temperature as measured by heat capacity is field dependent (see Figure S6 in SI), and 3) as stated above, the material adopts a different crystal structure below 9.7 K, with results in altered superexchange pathways. Notably, no transitions similar to the 9.7 K transition have been observed in other \ce{\textit{Ln}OF} compounds, with the exception of GdOF which shows a similar bump at 4.0 K, despite no change in structure was reported.\cite{Dutton2012} We note that the \ce{Tb^{3+}} and \ce{Gd^{3+}} ions both exhibit large magnetic moments. However, \ce{Gd^{3+}} ions are $L=0$ ions, resulting in very weak magnetoelastic coupling, whereas \ce{Tb^{3+}} ions possess strong quadrupolar and octupolar moments that couple efficiently to the lattice, leading to strong magnetoelastic coupling \cite{Vilarinho2022, Turrini2021}. We therefore hypothesize that the combination of strong magnetic moment and magnetoelastic coupling  of \ce{Tb^{3+}} makes TbOF stand out with respect to isostructural \ce{\textit{Ln}OF} compounds.

Upon cooling below 3.5 K, a set of well-defined magnetic reflections appears in the neutron diffraction pattern, consistent with the formation of a long-range magnetically ordered state. The magnetic reflections are in accordance with the propagation vector $\vec{k}=(0, 0.2569(1), 0.50)$ for the monoclinic unit cell, with $C\bar{1}$ symmetry. This propagation vector corresponds to a doubling of the crystallographic (monoclinic) unit cell along the \textit{c}-axis and an incommensurate modulation along the \textit{b}-axis, approximately quadrupling the magnetic periodicity along this direction. 

\begin{figure}[t]
    \centering
    \includegraphics{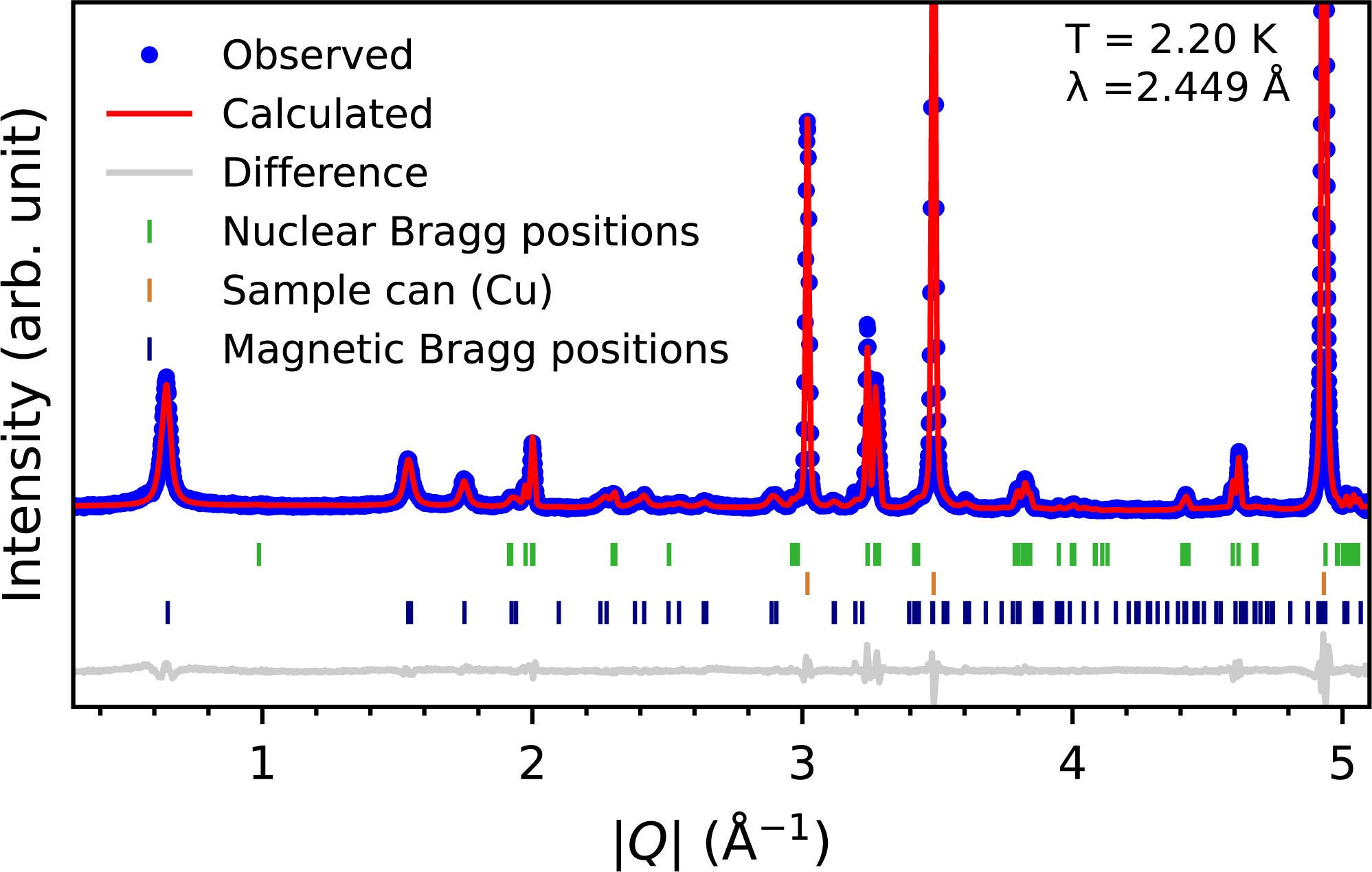}
    \caption{Rietveld refinement of the powder neutron diffraction data for \ce{TbOF} at 2.20 K ($\lambda = 2.449\;\text{Å}$). The nuclear and magnetic Bragg reflections are depicted by the green and blue tickmarks, respectively. The copper sample can gives rise to peaks indicated by the orange tickmarks and included in the refinement. The refinement was performed by fitting this data and that measured with $\lambda = 1.494 \;\text{Å}$ simultaneously using the same structural model (see Figure S7 in SI).}
    \label{fig:Rietveld 2.2K}
\end{figure}

\begin{figure}[ht]
    \centering
    \includegraphics{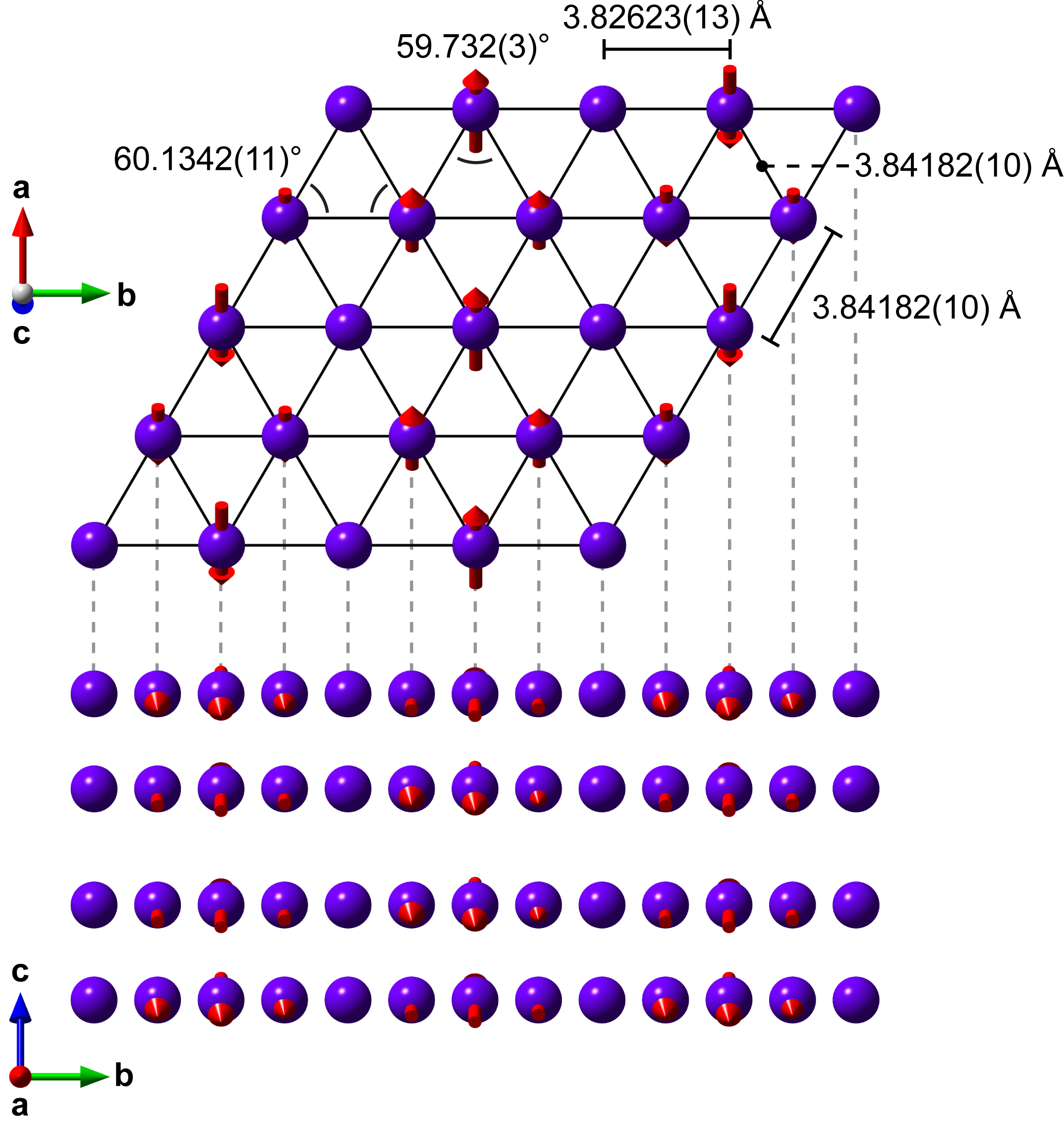}
    \caption{Visualization of the magnetic structure of \ce{TbOF} at 2.2 K, presented as a supercell. O and F are omitted for clarity. Top: a single Tb layer viewed perpendicular to the \textit{ab}-plane. The Tb–Tb distances and Tb–Tb–Tb angles are indicated, showing the isosceles nature of the triangles in the plane. 
    Bottom: view along the \textit{a}-axis. The dashed lines are included as a guide to the eye, connecting them to the top view.}
    \label{fig:MagneticStructure}
\end{figure}

For the magnetic order of the \ce{Tb^{3+}} ions, the symmetry analysis of the possible magnetic orders was performed with the BasIreps program embedded into the FullProf software.\cite{RodrguezCarvajal1993FULLPROF, basireps, RodriguezCarvajal2025} Having checked all the possible ordering possibilities, we come to the conclusion that the only possibility to describe the experimentally observed pattern of magnetic diffraction data is as a collinear spin density wave structure that incommensurately modulates along the crystallographic $b$-axis. The refinement is shown in Figure \ref{fig:Rietveld 2.2K} and the magnetic ordering is shown in Figure \ref{fig:MagneticStructure}.

As stated above, we note that the Tb triangles become isosceles during the phase transition to $\text{C}2/m$, resulting in significantly shorter Tb-Tb distances along the \textit{b}-direction (3.82623(13) compared to 3.84182(10) Å at 2.2 K). Therefore, the exchange coupling along this direction is stronger compared to the other edges of the triangular motif, resulting in the \textit{b}-direction being the modulation direction of the structure. Along the other Tb-Tb direction, however, a modulation with much larger periodicity is also observed, as shown in Figure \ref{fig:MagneticStructure}. The magnetic moments are slightly canted along the \textit{c}-axis, and reach an amplitude of $7.52\pm0.13$ $\upmu_{\text{B}}$.

In contrast to the $ab$-plane, wherein neighboring Tb atoms are connected via both oxide and fluoride bridges, along the $c$-axis, the Tb layers are alternately coupled by either oxide or fluoride bridges. As demonstrated by our DFT calculations using the room temperature TbOF crystal structure (discussed above), the Tb atoms connected via O are closer together than those connected via F. Furthermore, this inequivalence in bond length is significantly enhanced during the 9.7 K phase transition, as shown in Figure S5 in SI. As shown in \ref{fig:MagneticStructure}, short Tb-Tb distances are accompanied by antiferromagnetic interactions, while long Tb-Tb distances provide ferromagnetic interactions. Consequently, we observe an antiferromagnetic coupling between Tb atoms via O and a ferromagnetic coupling via F, resulting in a doubling of the magnetic unit cell along the $c$-axis. 

\begin{figure}[ht]
    \centering
    \includegraphics[width=0.4\textwidth]{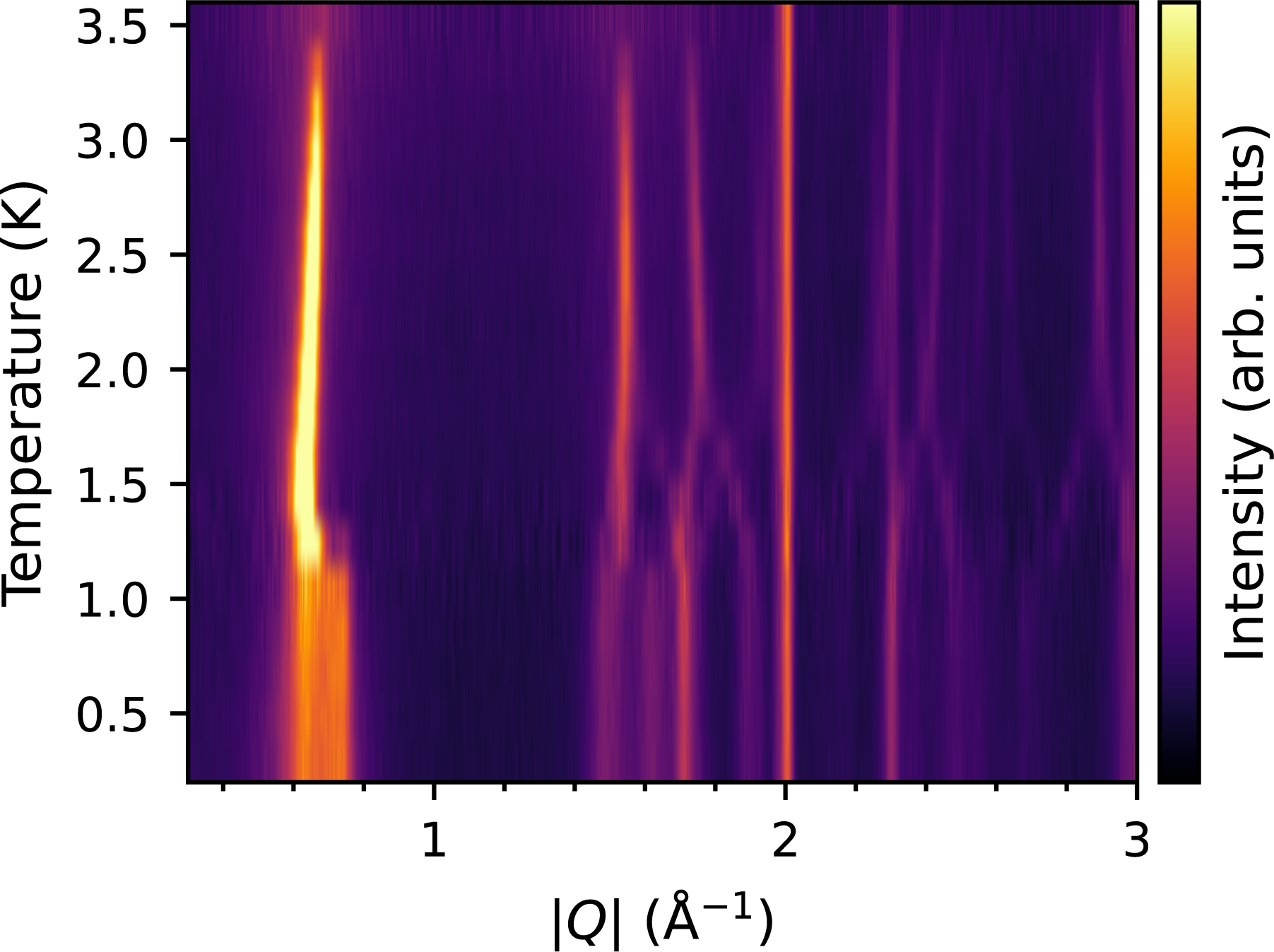}
    \caption{The temperature evolution of the neutron powder diffraction pattern of \ce{TbOF} from 0.2 K to 3.6 K in 0.2 K steps, collected using \(\lambda=1.886\;\text{Å}\). The diffraction patterns show the evolution of the magnetic structure, with the onset of magnetic peaks below 3.5 K and the magnetic structure adopting a different phase below 2.0 K, that settles below 1.0 K. Individual diffraction patterns are shown in Figure S8.}
    \label{fig:Heatmap HRPT}
\end{figure}

Notably, the magnetic behavior of \ce{TbOF} becomes significantly more complex upon further cooling. As shown by the temperature evolution of the neutron diffraction patterns in  Figure \ref{fig:Heatmap HRPT}, the above mentioned magnetic structure rapidly changes below 2.0 K. The transition occurs over a temperature range of around 1.0 K, wherein magnetic Bragg peaks shift or split, and additional peaks appear, before settling below 1.0 K. 

In order to gain insight into the nature of the settled magnetic state below 1.0 K, we focus on the lowest temperature data, obtained at 90 mK. We note that the magnetic phase could not be described by a single propagation vector, possibly due to the fact that multiple changes in the diffractogram appear at different temperatures. Instead we found two modulation vectors necessary to index all magnetic Bragg peaks present at 90 mK, as demonstrated by the preliminary Rietveld refinement shown in Figure S9. In our model, the first component retains the original modulation character, with $\vec{k_1}=(0, 0.2769(7), 0.50)$, indicating a slight change of the incommensurate wavevector along the \textit{b*}-direction. As a consequence, the ordered moment is reduced to $5.53\pm1.01\;\upmu_{\text{B}}$ and remains collinear with an enhanced canting of the magnetic moment along the \textit{c}-axis compared to 2.2 K.

A second magnetic component is required to account for additional magnetic intensity, which is characterized by a distinct incommensurate propagation vector $\vec{k_2}=(0.75, 0.7605(3), 0.50)$. The appearance of this second propagation vector signifies a change in the magnetic ordering, as was indicated by the symmetry analysis performed using the BasIreps program:\cite{basireps} whereas the original long-range order observed at 2.2 K involves spins confined to the \textit{ac}-plane, this vector introduces a transverse magnetic component that rotates the moments into the \textit{ab}-plane and imposes a different spatial periodicity along the \textit{a}- and \textit{b}-axis. This can be understood as a magnetic reorientation that appears only partially expressed.

While the combination of these two propagation vectors accounts for all magnetic Bragg peaks present at 90 mK, having checked all the possible ordering possibilities, the fit shown in Figure S9 appears to be the best possible fit. However, we note that the refinement is complicated by the complex magnetic behavior of \ce{TbOF}, as supported by the low-temperature magnetization data presented below.

\begin{figure}[ht]
    \centering
    \includegraphics{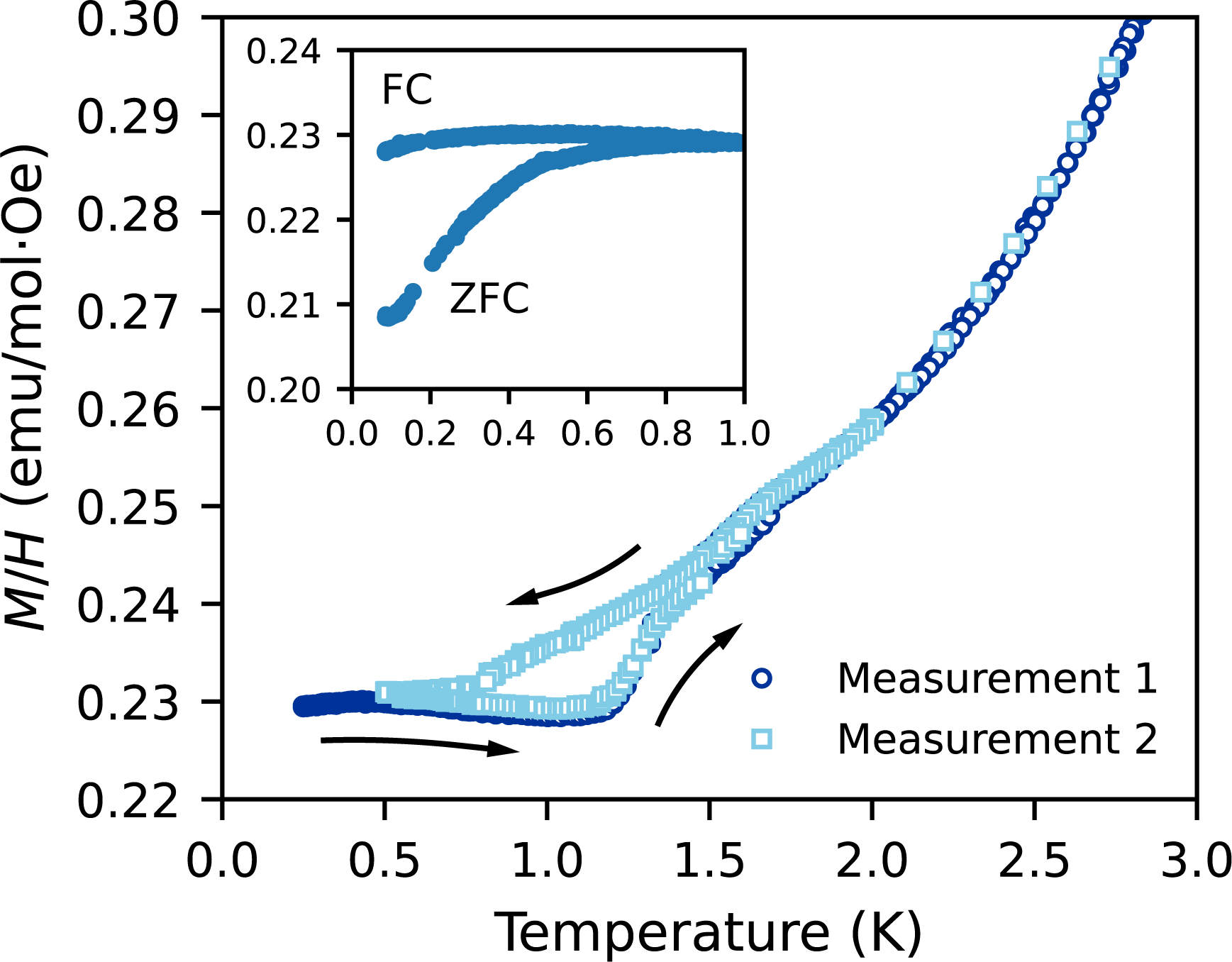}
    \caption{Temperature-dependence of the magnetization ($M/H$) for \ce{TbOF} at low temperatures. Two different measurements reproducibly display a hysteresis when measured while heating or cooling with $H=0.01\;\text{T}$. The inset shows additional measurements wherein a splitting between ZFC (zero-field cooled) and FC (field-cooled) protocols is observed below 0.6 K.}
    \label{fig:Low-T_SQUID}
\end{figure}

Most notably, upon further investigating the magnetism below 2.0 K, we find that the change in the neutron diffraction pattern coincides with a hysteresis in temperature-dependent magnetization, opening up a gap between the FC cooling and warming curves, as shown in Figure \ref{fig:Low-T_SQUID}. Furthermore, we observe a divergence between the FC and ZFC curves below 0.7 K, which hints at metastable behavior. Therefore, we argue that \ce{TbOF} has not fully reached the magnetic ground state at 90 mK. These two notable observations result from the geometric frustration inherent to the lattice, which is the origin of the complex magnetic behavior in TbOF.

\section{Conclusion} 
\noindent 
In conclusion, we have investigated the magnetic phase diagram of TbOF, a frustrated rhombohedral crystal system containing magnetic \ce{Tb^{3+}} ions, in great detail. Our findings establish TbOF as a rare example of a lanthanide oxyfluoride in which geometric frustration is paired with unconventional, temperature-dependent magnetic phases. We have demonstrated that, in contrast to earlier reports, \ce{TbOF} does not undergo a long-range antiferromagnetic ordering at 9.7 K, but rather undergoes a structural transition to a monoclinic phase that breaks three-fold rotational symmetry. In addition, this marks the onset of what appears to be short-range magnetic correlations at this temperature.

The system adopts three-dimensional long-range magnetic order below 3.5 K, harboring an incommensurate modulation in the Tb plane, with alternating ferromagnetic and antiferromagnetic interactions between the planes. Our DFT calculations confirm that these interactions occur via the longer Tb-F-Tb and the shorter Tb-O-Tb super-exchange pathways, respectively. Furthermore, we find the development of an additional, more magnetically complex phase at even lower temperatures, including metastable behavior and a unusual hysteresis effect in the magnetization. This rich magnetic phase diagram provides further insight into frustrated mixed-anion systems with magnetic $4f$ ions.

\section*{Acknowledgment}
\noindent This work is based on experiments performed at the Swiss spallation neutron source SINQ, Paul Scherrer Institute, Villigen, Switzerland. This project was supported by the Danish National Committee for Research Infrastructure through DanScatt and the ESS-Lighthouse Q-MAT. The computational resources (ænetone HPC) used in this work for the DFT calculations were supported by a Dutch Sector Plan grant awarded to NA. MEK acknowledges the research program \textit{Materials for the Quantum Age} (QuMat) for financial support. This program (registration number 024.005.006) is part of the Gravitation program financed by the Dutch Ministry of Education, Culture and Science (OCW).

\section*{Associated Content}
\subsection*{Supporting Information}

\noindent\textbf{Supporting Information Available:} crystallographic information files of TbOF measured at 14 K, 8 K, 2.2 K and 90 mK (CIF), and magnetic structure of TbOF measured at 2.2 K (mCIF). Additional X-ray diffraction, magnetization, temperature-dependent neutron powder diffraction, and magnetic field-dependent heat capacity measurements. Additional information first-principle DFT calculations and bond valence sum calculations (PDF). This material is available free of charge via the Internet at
\hyperlink{}{http://pubs.acs.org}.

\bibliography{References}

\end{document}


\include{authors}

\maketitle

\begin{figure}[ht]
    \centering
    \includegraphics{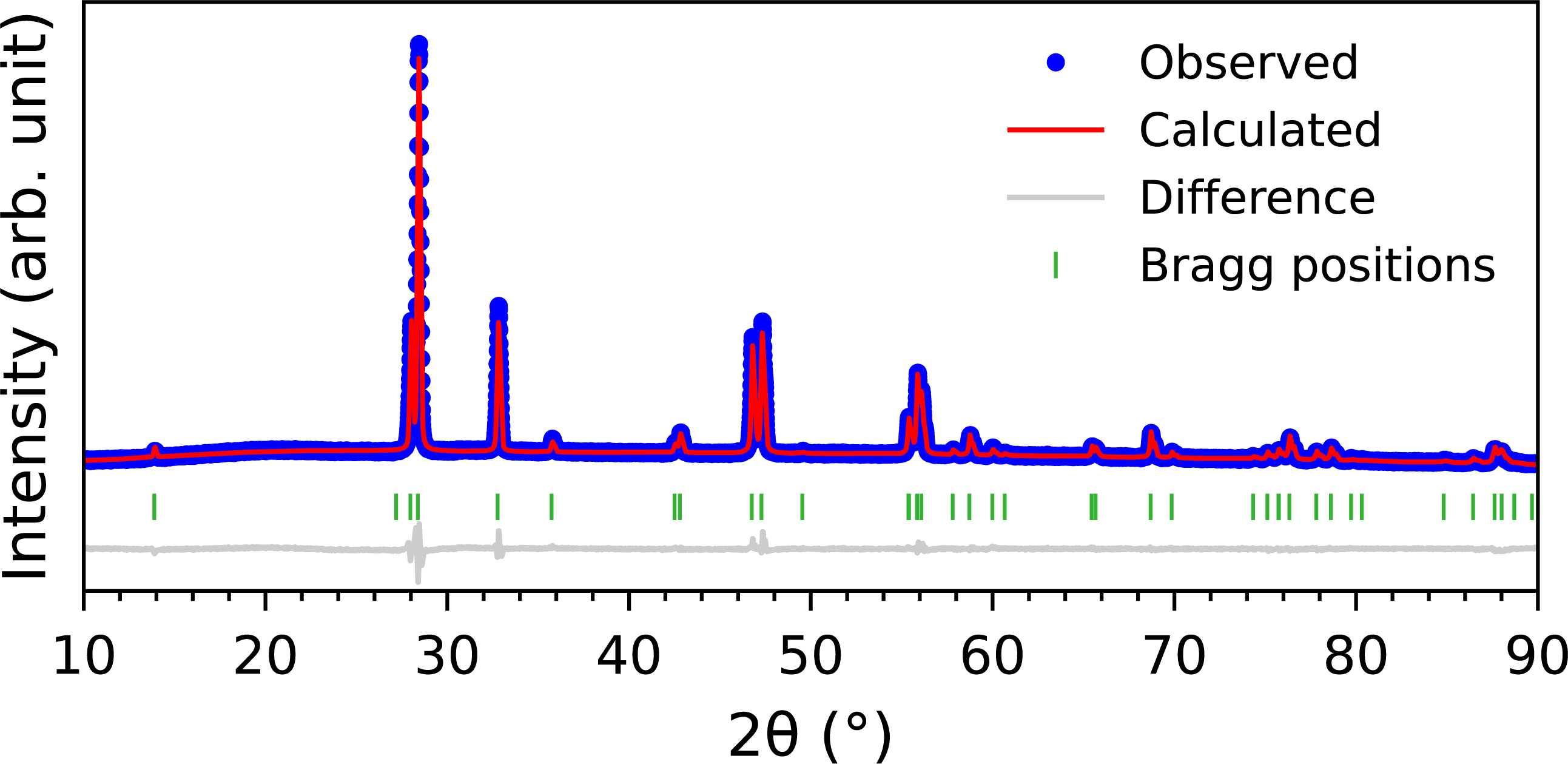}
    \caption{Powder X-ray diffraction pattern of \ce{TbOF} taken at room temperature, using CuK\(\alpha\) radiation.}
    \label{fig:Powder XRD}
\end{figure}

\begin{figure}
    \centering
    \includegraphics{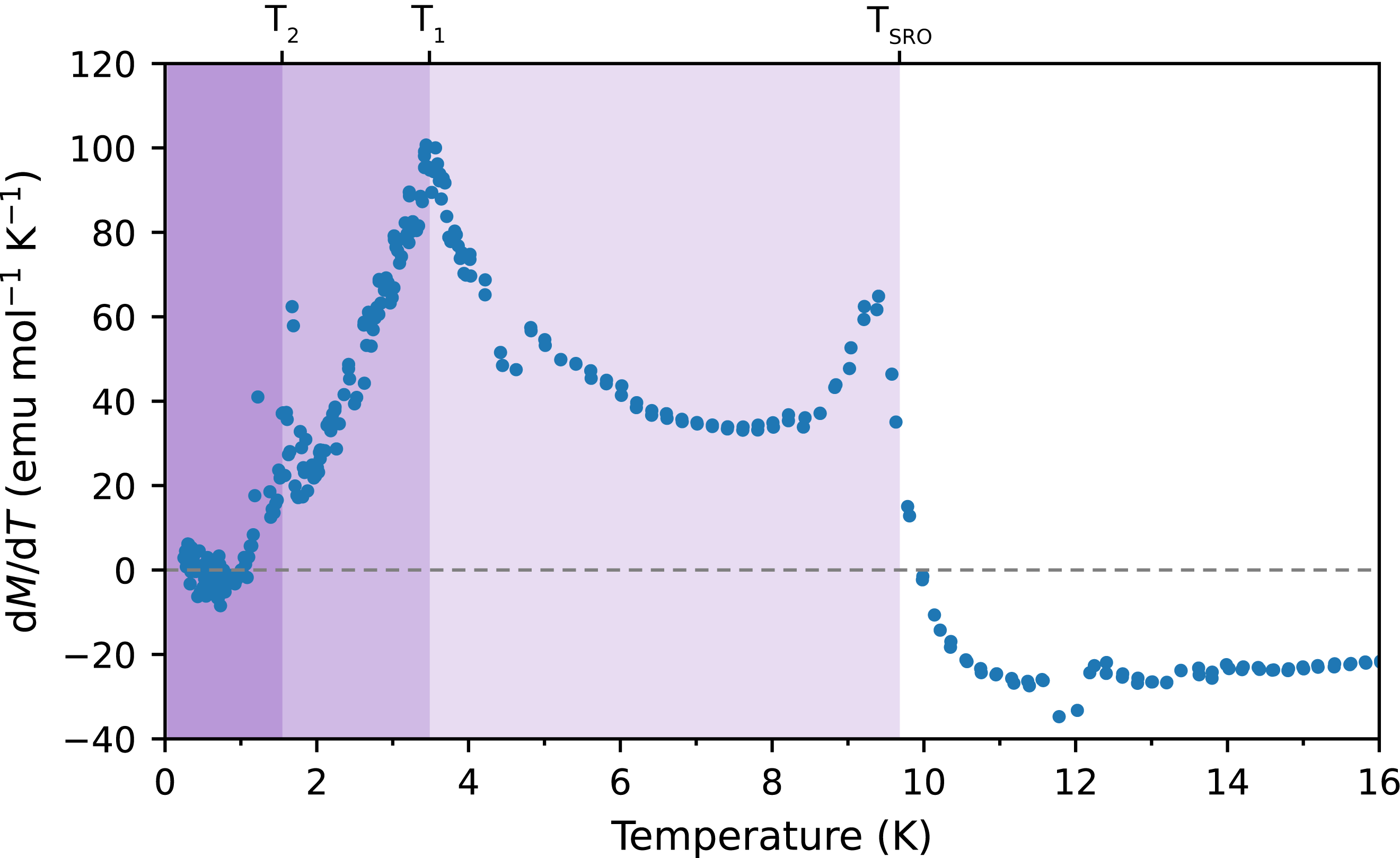}
    \caption{The temperature-dependence of the first derivative of the magnetization ($\frac{\text{d}{M}}{\text{d}T}$) for \ce{TbOF}. 
    The magnetization data is a compilation of measurements above and below 1.8 K, measured under applied magnetic fields of 0.1 and 0.01 T, respectively. $T_1$, $T_2$ and $T_{\text{RSO}}$, represent the two magnetic order and short-range-order transitions as determined by the specific heat data presented in Figure 2 in the main text.
    }
    \label{fig:Susceptibility derivative}
\end{figure}

\begin{figure}
    \centering
    \includegraphics{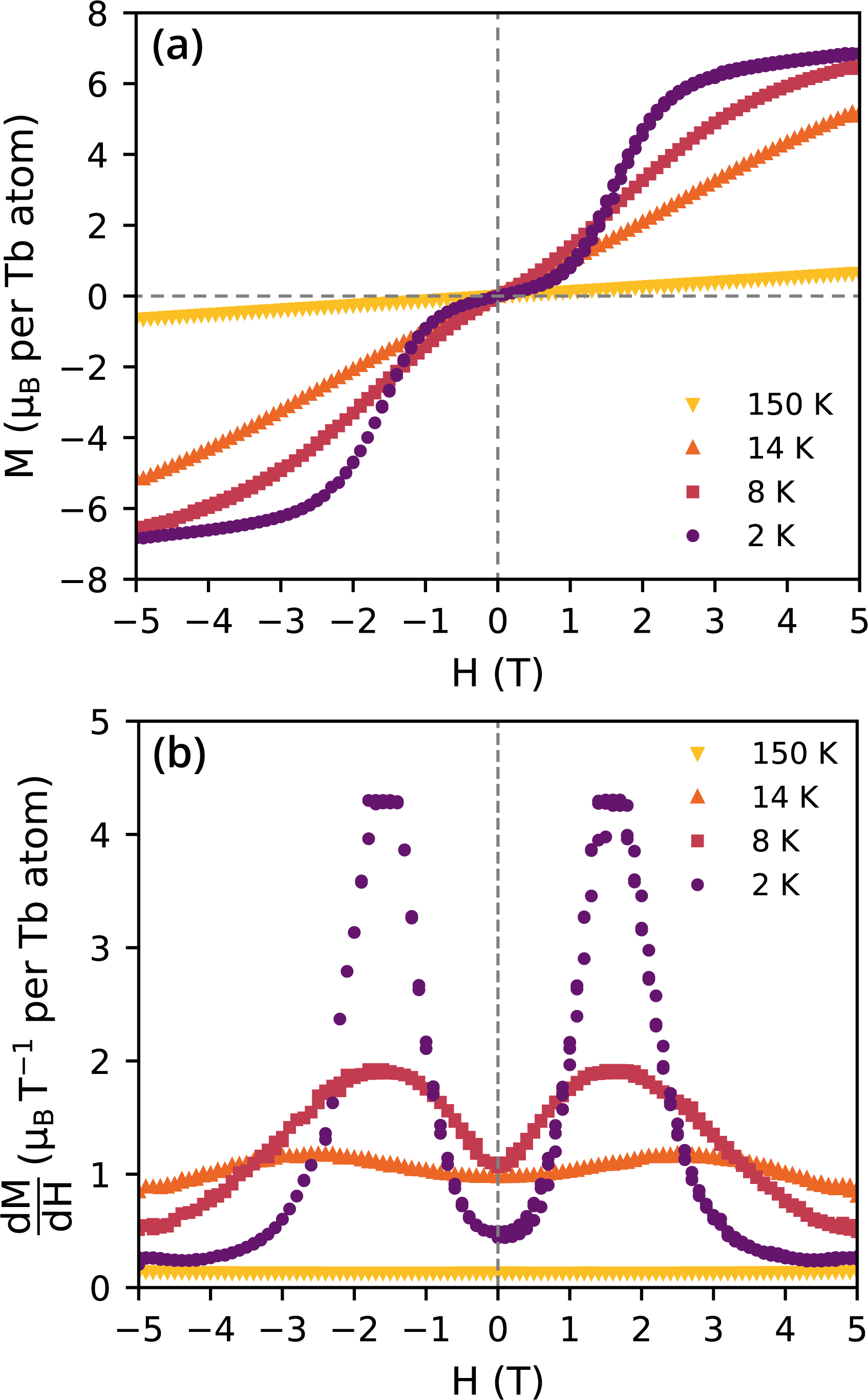}
    \caption{(a) Isothermal magnetization of \ce{TbOF} performed at T = 2.0 K, 8.0 K, 14.0 K, and 150.0 K, with the applied field $H$ varied from $-5.0\;\text{T}$ to $+5.0\;\text{T}$. (b) The derivative of the magnetization with respect to the applied magnetic field, $\text{d}M/\text{d}H$.}
    \label{fig:MvsH}
\end{figure}

\begin{figure}[ht]
    \centering
    \includegraphics{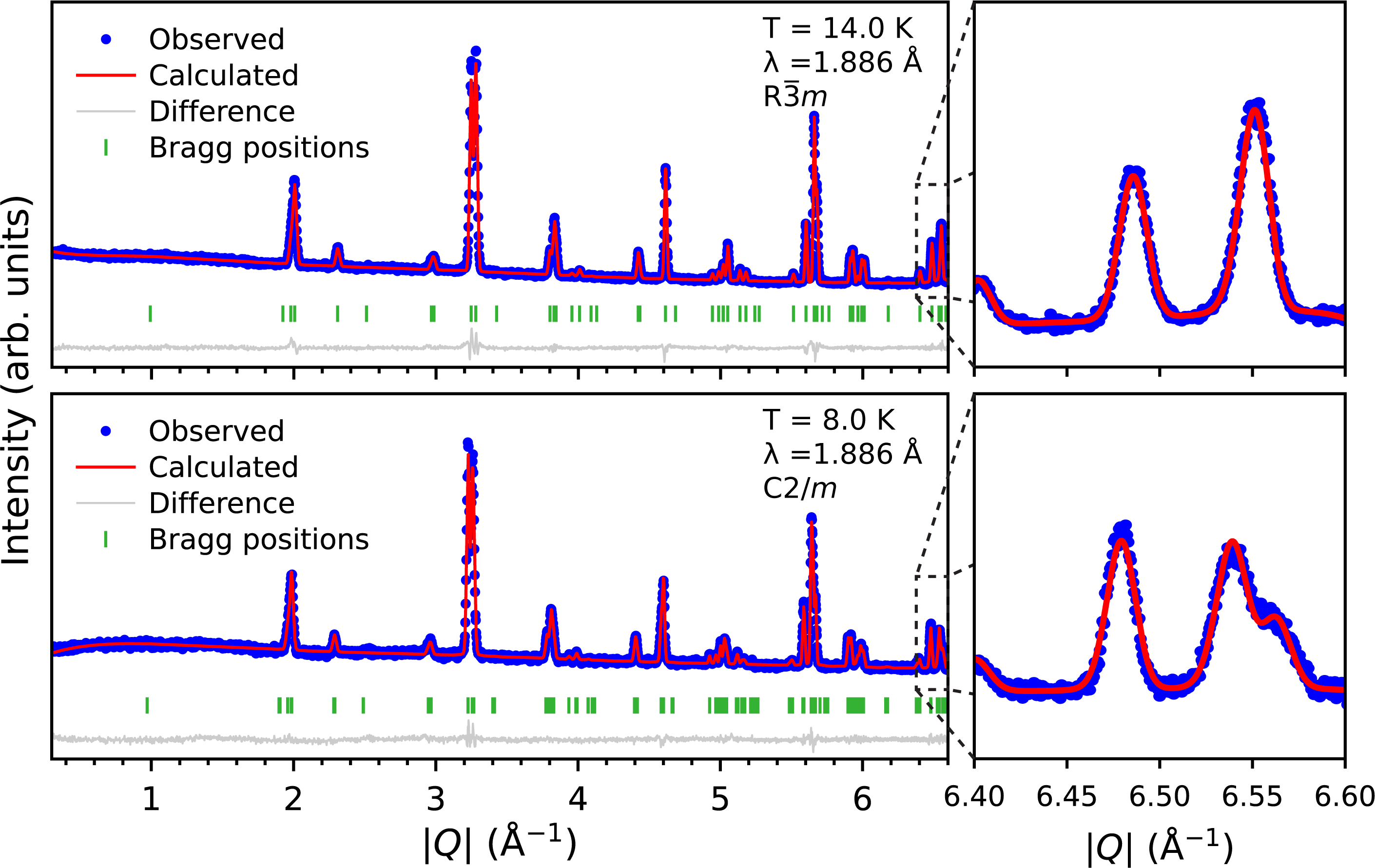}
    \caption{Rietveld refinement of neutron powder diffraction patterns of \ce{TbOF} collected at 14.0 K (top) and 8.0 K (bottom) collected using \(\lambda=1.886\;\text{Å}\). At 14.0 K \ce{TbOF} adopts a rhombohedral \(\text{R}\Bar{3}m\) (No. 166) structure. At 8.0 K \ce{TbOF} has undergone a phase transition to a monoclinic \(\text{C}2/m\) (No. 12) structure.}
    \label{fig:Rietveld, Nuclear change}
\end{figure}

\begin{figure}[ht]
    \centering
    \includegraphics{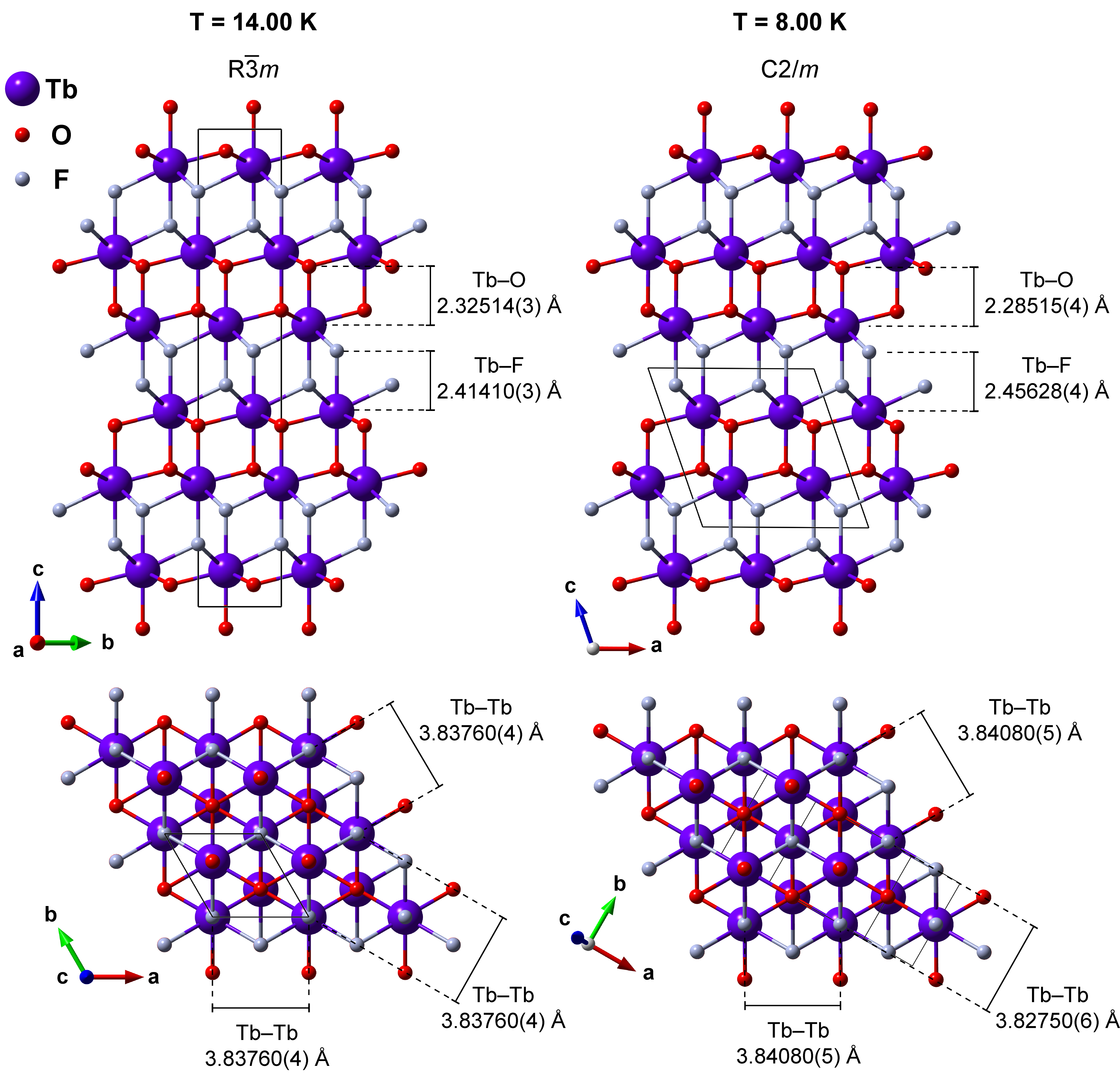}
    \caption{Crystal structures of TbOF at 14 K (left) and 8 K (right), viewed along different crystallographic axes. Unit cells and relevant bond lengths are indicated in both structures. }
    \label{fig:Nuclear phases comparison}
\end{figure}

\begin{figure}[ht]
    \centering
    \includegraphics{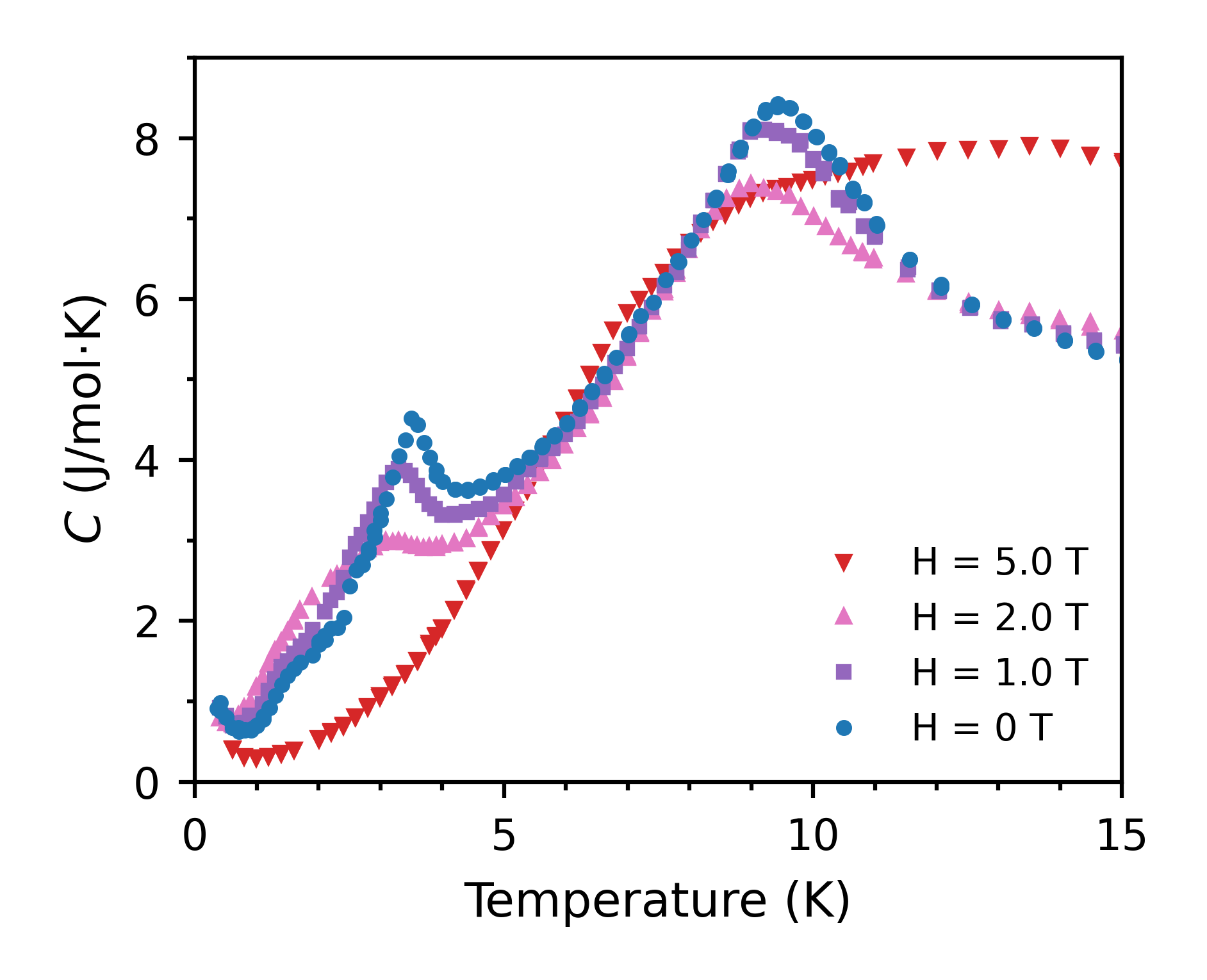}
    \caption{Heat capacity measurements for \ce{TbOF} under various static applied magnetic fields. 
     }
    \label{fig:Specific Heat under field}
\end{figure}

\begin{figure}[b]
    \centering
    \includegraphics{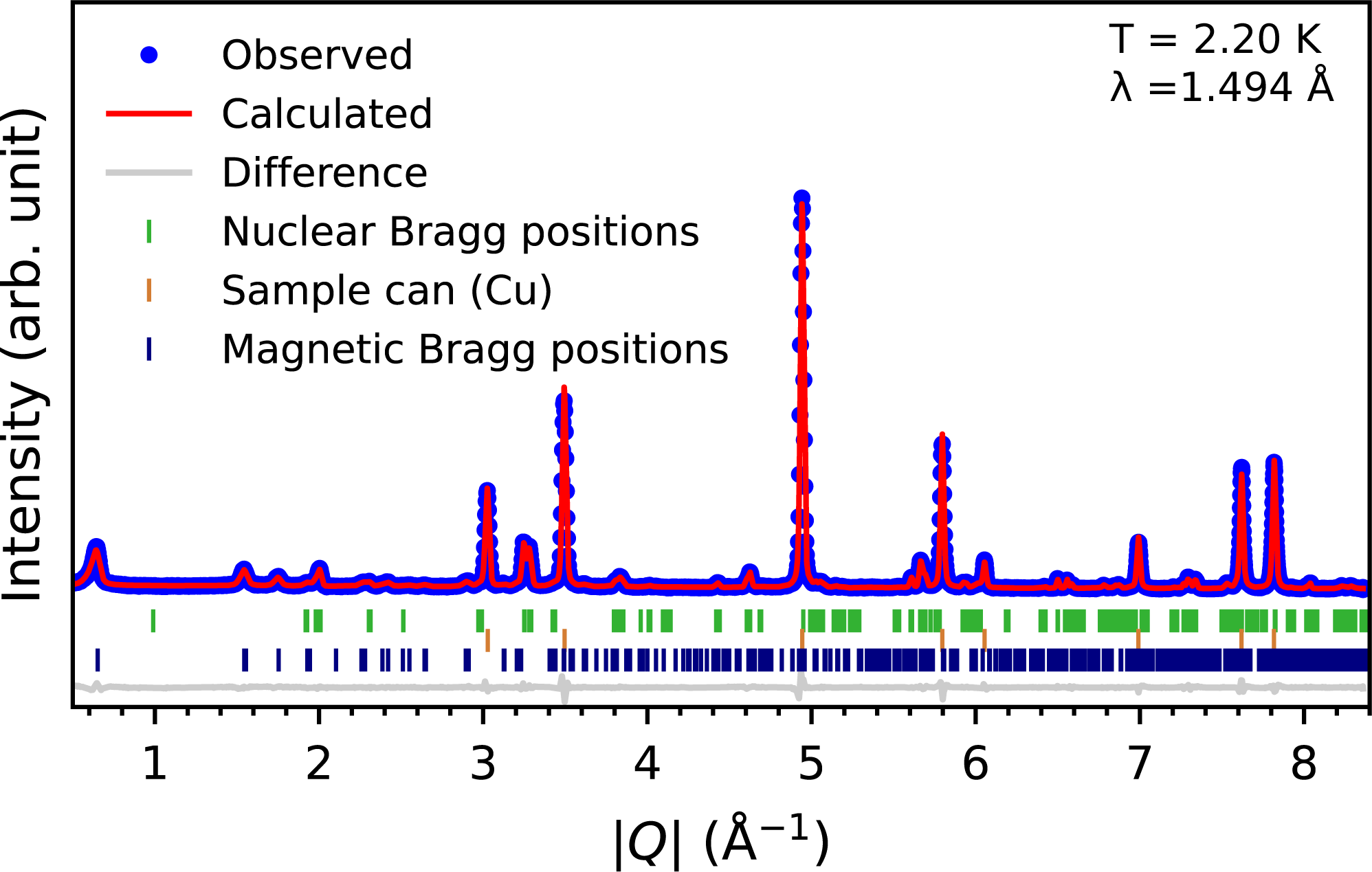}
    \caption{Rietveld refinement of the powder neutron diffraction data for \ce{TbOF} at 2.20 K ($\lambda = 1.494\;\text{Å}$). The nuclear and magnetic Bragg reflections are depicted by the green and blue tickmarks, respectively. The copper sample can gives rise to peaks indicated by the orange tickmarks and were included in the refinement. The refinement was performed by fitting this data and that measured with $\lambda = 2.449 \;\text{Å}$ simultaneously using the same structural model (see Figure 4 main text). 
    }
    \label{fig:Rietveld 2.2K Supplement}
\end{figure}

\begin{figure}[ht]
    \centering\includegraphics{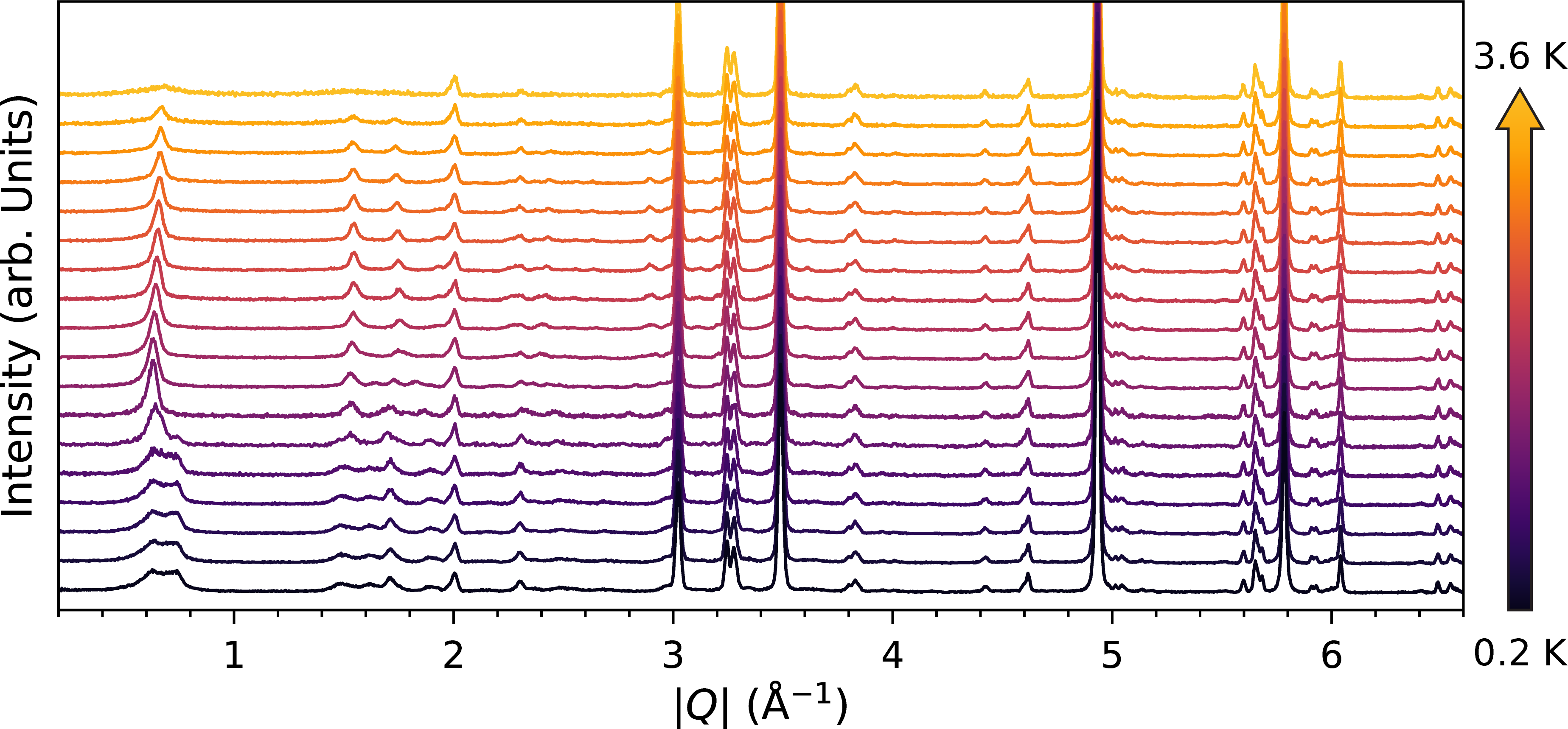}
    \caption{Temperature evolution of the neutron powder diffraction pattern of \ce{TbOF} from 0.2 K to 3.6 K, collected using \(\lambda=1.886\;\text{Å}\), shown with vertical offset for clarity. The diffraction patterns show the evolution of the magnetic structure, with the onset of magnetic peaks below 3.5 K and the magnetic structure adopting a different phase below 2.0 K, that settles below 1.0 K. The same data is presented as a heatmap in Figure 6 of the main text.}
    \label{fig:Magnetic_Phase_Change}
\end{figure}

\begin{figure}
    \centering
    \includegraphics{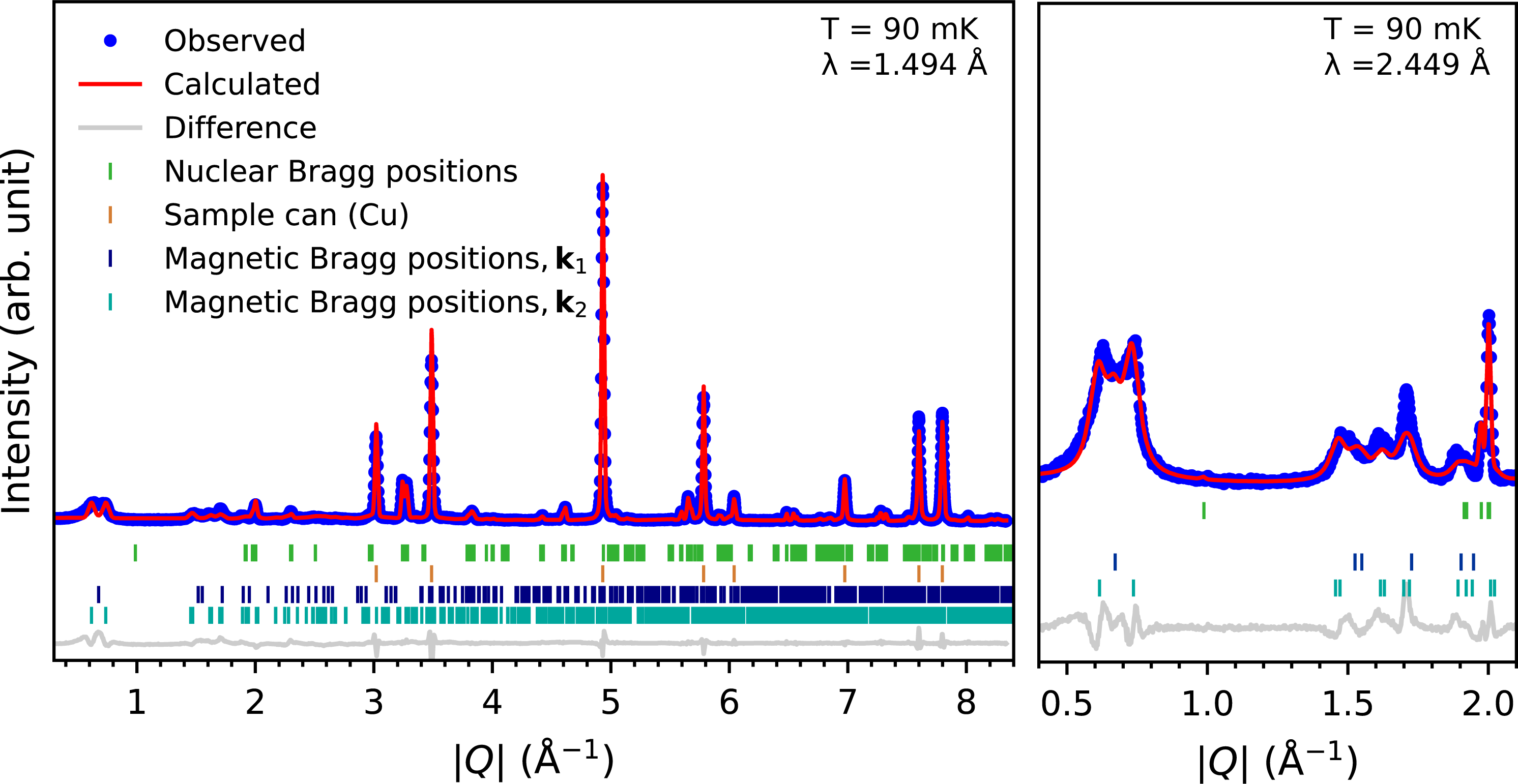}
    \caption{Rietveld refinement of the powder neutron diffraction data for \ce{TbOF} at 90 mK ($\lambda = 1.494\;\text{Å}$). The nuclear Bragg reflections are depicted by the green tickmarks. The magnetic reflections for $\vec{k_1}=(0, 0.2769(7), 0.50)$ and $\vec{k_2}=(0.75, 0.7605(3), 0.50)$ by the blue and turquoise tickmarks, respectively. The copper sample can gives rise to peaks indicated by the orange tickmarks and were included in the refinement. The right box is a zoom in of the low Q data, measured with $\lambda = 2.449\;\text{Å}$ and fitted with the same model.}
    \label{fig:Rietveld 90mK}
\end{figure}

\clearpage

\section*{First-principle DFT calculations}
Two distinct structural configurations of \ce{TbOF} were studied: \textbf{config1}, with short Tb–O bonds and long Tb–F bonds, and \textbf{config2}, with short Tb–F bonds and long Tb–O bonds, see Table S1. \\

\begin{table}[ht]
\centering
\caption{Crystallographic parameters for TbOF at 300 K, obtained from Rietveld refinement of powder X-ray diffraction data, used in the calculation. Atom positions given in $x$, $y$, $z$ fractional coordinates.}
\begin{tabular}{|l|l|l|}
 \hline
                        & \textbf{config1} & \textbf{config2}  \\
 \hline \hline
crystal system          & Trigonal    & Trigonal    \\
space group             & $\text{R}\bar{3}m$    & $\text{R}\bar{3}m$    \\
Z                       & 6    & 6    \\
$a$ (Å)                   & 3.8411(4)  & 3.8411(4)  \\
$b$ (Å)                   & 3.8411(4)  & 3.8411(4)  \\
$c$ (Å)                   & 19.1247(8)  & 19.1247(8) \\
$\alpha$ ($\degree$)    & 90    & 90    \\
$\beta$ ($\degree$)     & 90    & 90    \\
$\gamma$ ($\degree$)    & 120    & 120    \\
volume ($\text{Å}^3$)   & 244.368(5)    & 244.368(5)    \\
Tb ($x$,$y$,$z$) & 0.00000    & 0.00000   \\
 & 0.00000    & 0.00000   \\
 & 0.74210    & 0.74210   \\
O ($x$,$y$,$z$) & 0.00000    & 0.00000   \\
 & 0.00000    & 0.00000   \\
 & 0.86900   & 0.61780 \\
 F ($x$,$y$,$z$) & 0.00000    & 0.00000   \\
 & 0.00000    & 0.00000   \\
 & 0.61780   & 0.86900  \\
  \hline
\end{tabular} \label{Table: Nuclear phases}
\end{table}

\noindent The relative stability of the two configurations was evaluated for both the primitive unit cell ($(1\times1\times1)$) and for a $2\times2\times1$ supercell to confirm that the energetic ordering is not affected by finite-size effects. All energies are reported as relative energies referenced to config1.

\begin{equation*}
    \Delta E = E_\text{config}-E_\text{minimum}.
\end{equation*}
For the primitive unit cell, the energy difference between the two orderings is $\Delta E = E_{\text{config2}}-E_{\text{config1}}=14.1556\;\text{eV}$ (config2 is higher by 14.16 eV). For the $2\times2\times1$ supercell the corresponding energy is $\Delta E = E_{\text{config2}}-E_{\text{config1}}=45.9221\;\text{eV}$ (config2 is higher by 45.92 eV). The energetic ordering is preserved upon cell-size enlargement, indicating the relative stability is robust to finite-size effects.

At the static DFT level, config1 is significantly more stable than config2. For the primitive unit cell containing 18 atoms, config2 is higher in energy by 14.16 eV per unit cell, \textit{i.e.} 0.79 eV per atom. The same ordering is obtained for the $2\times2\times1$ supercell (72 atoms), where config2 is higher in energy by 45.92 eV per supercell, equivalent to 11.48 eV per primitive unit cell or 0.64 eV per atom after normalization. The agreement between different cell sizes shows that the stability of config1 is intrinsic and not caused by finite size effects.

To examine finite-temperature stability, \textit{ab initio} molecular dynamics (AIMD) simulations were performed at 20 K using a time step of 1 fs. config1 equilibrates rapidly and shows small and stable energy fluctuations, indicating a dynamically stable structure. config2 exhibits higher energy at early simulation times before relaxing towards lower energy configurations similar to config1, indicating a less stable initial structure.

Both static DFT calculations and AIMD simulations show that config1 is energetically more stable than config2.

\clearpage

\section*{Bond valence sum calculation}
In addition to the DFT calculations above, we have also calculated the bond valance sum (BVS) for both configurations, using the bond distances from the two configurations listed in Table S1 and applying the formula reported by Brown and Altermatt:\cite{Brown1985} 

\begin{equation*}
    s = \text{exp}[(r_0 - r)/ B]
\end{equation*}

\noindent The parameters used were taken from the paper by Brese and O'Keeffe;\cite{Brese1991} with $r_0 =$ 2.049 Å and 1.936 Å for the Tb–O bond and the Tb–F bond, respectively, and $B = 0.37$ Å, as in line with Brown and Altermatt \cite{Brown1985}.

The results are given in Table S2. Due to the expected 3+ valence of Tb, config1, with short Tb-O and long Tb-F bonds is the most plausible structure, in agreement with our DFT calculations.\\

\begin{table}[]
\caption{Bond valence sum (BVS) calculations for both configurations given in Table S1.}
\begin{tabular}{|l|ll|l|l|}
\hline
Configuration  & \multicolumn{2}{l|}{Bond} & Bond length (Å) & BVS \\
\hline \hline
\textbf{Config1} & Tb–O      & $\times1$        & 2.377(6)       & 0.412(7)    \\
         & Tb–O      & $\times3$        & 2.2751(13)     & 0.543(2)    \\
         & Tb–F      & $\times1$        & 2.427(4)       & 0.265(3)   \\
         & Tb–F      & $\times3$        & 2.4591(17)     & 0.243(1)   \\ \hline
         & \textbf{Total}          &    &                & \textbf{3.035(8)} \\
\hline \hline
\textbf{Config2} & Tb–O      & $\times1$        & 2.427(4)       & 0.340(4)     \\
         & Tb–O      & $\times3$        & 2.4591(17)     & 0.315(1)    \\
         & Tb–F      & $\times1$        & 2.377(6)       & 0.423(7)    \\
         & Tb–F      & $\times3$        & 2.2751(13)     & 0.561(2)   \\ \hline
         & \textbf{Total}          &    &                & \textbf{2.854(9)} \\
\hline
\end{tabular}
\end{table}

\clearpage

\bibliography{References}